\documentstyle[12pt]{article}

\newcommand{\be}{\begin{equation}}
\newcommand{\ee}{\end{equation}}

\newcommand{\bary}{\begin{eqnarray}}
\newcommand{\eary}{\end{eqnarray}}
\begin{document}
\title{The Neutron as an  Alternative Explanation for the
     Trans GZK Cosmic Rays}
{\author{
Adolfo De Un\'anue,
Sarira Sahu,
and
Daniel Sudarsky\\
Instituto de Ciencias Nucleares,\\
Universidad Nacional Autonoma de Mexico,\\
Circuito Exterior C.U., A. Postal 70-543,
04510 Mexico D.F.
}}
\date{ }
\maketitle
\begin{abstract}
\noindent
We consider the possibility that the Ultra High
Energy Cosmic Rays arriving
to Earth  might be neutrons instead of protons. We stress that in
such case the argument for the GZK cutoff is weaker and that it is
conceivable that neutrons would not be affected by it. This scenario
would require the neutron to start  with an energy larger than the observed
one, in order to be able to travel the distances involved, within its proper life-time.
It  must then  loose  most of the extra energy through interaction  with the
galactic dark matter  or some other matter in the intergalactic medium.
\end{abstract}
\vfill
\eject

\section{Introduction}
Cosmic rays have often been the source of great insights into the workings of
the cosmos and the fundamental laws of physics. The most energetic of these
cosmic rays are thought to be ultra high energy protons, originating at
distances of more than $100$ Mpc. The distribution of the particle showers,
which are probably hadronic, has not indicated so far any appreciable
anisotropy, although the statistics for these high energy events is rather poor
so far. Fortunately, the Auger project is expected to clarify the situation
soon  (its initial results do not seem to offer a conclusive
answer$^1$
). There is however an almost inescapable expectation that
there would be a sharp cutoff in the energies of these showers, as the highly
energetic primary protons would interact with the CMB photons loosing
most of their energies, until they drop below $\sim 3\times10^{19}$ eV, the so
called GZK cutoff$^{2, 3}$. 
This conclusion relies essentially on  the
 { \it measured}  proton  energy loss and cross section
for the $p+\gamma_{CMB}\rightarrow \Delta^{0}\rightarrow p+ q_{\pi} \pi $
processes,  the  {\it well established} thermal distribution of the CMB photons
with $ T \approx 2.7K^{o}$  and the Lorentz invariance which allows us to
connect the rest frame of the UHECR with the laboratory frame where
measurements of  the above mentioned cross sections are carried out. The
observation during the last years of a
series of events with energies quite above the GZK barrier, has produced a lot
of excitement in various communities. First and foremost, in the community of
cosmic ray physicists, but also in a sector of the community working on tests
of
fundamental symmetries and/or in quantum gravity. In fact the last few years
have witnessed the resurgence of interest into the possible existence of a
violation of strict Lorentz invariance associated with the microstructure of
the
space-time fabric, a type of ether-like feature in the Universe, presumably
connected to the quantum mechanical aspects of gravitation. As it turns out the
above mentioned absence of the GZK cutoff has played a central role as a
motivation for much of the ongoing excitement about the possibility of
violation
or modification of exact Lorentz Symmetry$^{4, 5, 6, 7, 8}$. 
These proposals face severe
problems.  In the first case, where one assumes that Lorentz Invarance is
broken
by the existence of a preferential rest frame tied somehow to the nature of
space time associated with Quantum Gravity, the problem resides in the fact
that
although one might initially postulate that the effects would be only
noticeable
at very high energies, the appearance of arbitrarily high energy particles in
the radiative corrections to any physical process, would transfer the influence
of the privileged frame to low energy processes where the effects would have
been observed long ago$^9$. 
In the other schemes where one hopes to
deform the representation of the Lorentz algebra by some nonlinear terms, the
problem is that one ends with an scheme in which the addition of our momenta
depends of the ordering of the particles, the existence of high energy
particles
even in remote regions of the universe would affect the local processes, or one
would rule out bodies with masses larger than the Planck mass.  All these
problems are of course so serious that one can not hope (despite the efforts of
some of its proponents) to use these schemes to address any phenomenological
issue, at least until those problems are resolved. For a more extensive
discussion of these issues see Ref.[10, 11, 12, 13, 14, 15]. 
Furthermore, in view of
the radical nature of any proposal challenging the validity of Special
Relativity, and the serious problems encountered by the existing proposals it
seems worthwhile to search for less exotic alternatives to understand the
apparent absence of GZK cutoff. It is clear that all proposals to address the 
problem at hand  will be "exotic" to some degree ( otherwise there would not be
a problem), thus we must emphasize that the criteria of exoticity in this
context must be taken as a relative one.  This relative scale is in this case 
very high and is represented by the proposals that challenge the validity of
Special Relativity, which as we have indicated face enormous (perhaps
insurmountable) obstacles.

The observation of such ultra high energy cosmic rays poses, in fact, two
challenges: What is exactly the mechanism that accelerates them to those
energies? and, how do they manage to travel beyond 50 Mpc despite the
expectation of a GZK cutoff?  The first issue represents a challenge mainly to
astrophysics, because the Hilles criteria$^{16}$ 
indicates that the
simplest acceleration mechanisms could not produce particles with the desired
energies in systems that are known to be within a reasonable distace.  The
second issue represents a very serious challenge to fundamental physics as even
if one could find a mechanism to produce the highly energetic particles, the
well established low energy physics of proton photon interaction together with
the ubiquitous cosmic microwave photon distribution should prevent such
particles from traveling the required distances without loosing most of their
energy.  The critical nature of this challenge can be noted by the apparent
willingness of the physics community to consider doing away with so cherished,
basic and well established principles as those underlying Spatial Relativity.
This article will thus mainly address the second issue while the first one,
regarded as essentially an astrophysical puzzle to be tackled  independently,
will only be touch upon in order to describe some alternatives that have been
considered so far.

There have been of course many proposals to deal with this problem, all of them
involving new physics in some way or another.  The fact that the observed
energies of the UHECRs pose also a serious challenge to the acceleration
mechanism in conventional astrophysical objects, explains the enhanced
atractiveness of proposals that involve new energy sources for the 
production of UHECRs$^{17}$. 
For instance, it has been suggested that, the UHECRs can be
produced in the decay of very massive ($m_X\ge 10^{12}$ GeV) and long lived X
particles, originating from high-energy processes in the early
Universe$^{17,18}$, 
the so called ``top down model''.  However, these
particles must decay within one GZK interaction length to avoid energy loss. At
the same time, their density must be large enough to give detectable flux of
UHECRs. A potential problem with this scenario is that, it predicts higher flux
of UHE photons than protons. Another recent  approach proposes that a strong
galactic halo sized magnetic field as a mechanism for isotropization of the
flux of UHECR of nearby origin$^{19}$, 
but unfortunately this scenario poses serious  astrophysical 
problems$^{20}$. 

  We will focus here on the possibility that the
ultra high energy cosmic rays are neutrons instead of protons. There are
several
reasons for doing this. First, the neutron is almost identical to the proton as
far as its hadronic properties and, thus, the particle showers produced by it
would be very similar to those produced by a proton. Second, one of the
possible sources of UHECR are compact object mergers, and these objects have
neutrons as their major constituent.  Third, despite what might appear at first
sight, essentially all astrophysical mechanisms that produce high energy
protons, are equally good sources for high energy neutrons; the reason is, of
course, the inverse beta decay of the energetic protons in interaction with the
electrons in the astrophysical source.  The problem is of course the mechanism
of acceleration resulting in such highly energetic particles. As we have said
our point of view here will be to address the fundamental physics issues and
regard this last but important issue to be an essentially  astrophysical
problem to be touched upon only
superficially. 
Finally, there is almost no direct experimental data of the way neutrons
interact with photons (or other particles) in the energy regime of 
interest$^{21,22}$, 
and thus, any suggestion of a conflict with
special relativity would be based on theoretical arguments which could in turn
be subject to questioning. In fact, the neutral character of the neutron
suggests, \emph{a priori}, that its interaction with the photons of the cosmic
microwave background would be substantially reduced as compared with that of
the
proton.  On the other hand, one could estimate the energy loss and cross
section
for the $n+\gamma_{CMB}\rightarrow \Delta^{0}\rightarrow n+q_{\pi}\pi $
process,
and
get a neutron mean free path similar to that of the proton. However, we should
keep in mind that this analysis does not rest on unshakable grounds .  In fact,
the strongest theoretical arguments on which these sort of calculation relies
on, are the phenomenological models known as \emph{Vector Meson Dominance}
(VMD)
or its variants such as \emph{Photon Hadronic Structure Approaches} (PHSA),
which are themselves not Lorentz covariant$^{23}$.
Thus, it seems  a logical possibility, that in the search for the "relatively
least exotic options", one should 
 question
such theoretical predictions if they are not corroborated by independent
empirical
evidence.  The point is in fact, not only that questioning the Lorentz
Violating
VMD seems far less radical than questioning Special Relativity, but also that
it
would be logically inconsistent to question Special Relativity, based, in part,
on the use of a Lorentz Violating postulate.

We emphasize that we do not challenge, in principle, the standard model as the
correct theory to treat the neutron's interaction with other known
particles. However, at the energies of interest QCD, is not necessarily in its
perturbative regime and the procedures to do calculations are not fully
reliable. In other words, we do not have a fully proven algorithm to calculate
the neutron cross-sections with different particles at these energies.  
  We must recall that the crux of the argument for the GZK cutoff for protons
relies in the well established
  pion photo-production cross section at the center of mass energy of the
$\Delta $ resonance,  a reaction in which the high energy proton looses an
important part of its energy ( of  order $20\%$ per reaction). Above this
threshold the cross section decreases rapidly  by a factor of $7$. This is the
regime where we question the identification of the measured characteristics of
the proton-photon interactions with those corresponding  to photon-neutron.
  At higher  neutron energies $E$  the center of mass energy $s$  becomes
  larger  (recall that $s$ grows like $ E^{1/2} $), eventually reaching the
  point where perturbative QCD calculation should be reliable, and the neutron
  and proton cross section and energy losses via pion photo production or
  other channels should become nearly  identical ( except  for a factor $2/3$
  corresponding  due to the difference in the sum  quarks charges
  squares). This should happen when the center of mass energy for the photon-
  quark ( or  photon-parton) system reaches the several $ 100 MeV$ level,
  corresponding to  a neutron energy  of about $E \approx  10^{21}  eV$
  depending on the exact percentage of the neutron energy carried by the
  corresponding parton. However then  the cross section should be  essentially
  $2/3$ of the corresponding proton cross section  which is a factor of about
  $ 7$ smaller than at the $\Delta$ resonance$^{24}$ 
(so between the two  effects one should get an full order of magnitude
decrease in the cross section),  while the mean  energy loss per reaction
remain  very similar and certainly  do not  change enough to compensate
for the above factors$^{25}$.
Thus   it is unclear whether  this would be a problem for the model.

In order
to make a reasonable lower bound estimate of the energy loss of neutrons with
the CMB we will use only the phenomenological Pauli interaction characterizing
the nucleon's magnetic moment coupling to photons.  On the other hand,
regarding
the nucleon interaction with the exotic particles, the situation is a total
unknown therefore    we will treat this part at a purely phenomenological
level.

The obvious problem that our proposal would encounter is that even at energies
like $10^{20} eV$ a neutron would, due to its finite lifetime, manage to
transverse only a fraction of the $100$ Mpc that separate us from the likely
sources of these energetic particles. The obvious solution is to assume that
the
neutron is in fact substantially more energetic, thus allowing it to cover the
distance involved, and, that it looses most of its energy, due to interactions
with matter during its trip toward us, so that it arrives to Earth with the
$10^{20}$ eV of the observed showers.

We will be considering in fact two scenarios: In the first, the neutron looses
its ``excess'' energy very slowly during
its whole intergalactic trip, through its interaction with some unknown
component of the intergalactic media.  In
the second scenario the neutron travels
through the intergalactic media, interacting only with the CMB of photons and
neutrinos, and loosing almost no energy, while at the end of the trip it losses
its ``excess'' energy interacting with the dark matter in the galactic halo.

\section{General Requirements}\label{general_requirements}
The requirements on the models are then twofold: That the neutron travels from
a
source at $D_S =10^2 $ Mpc and arrives to Earth with an energy of $10^{20}$ eV,
and, that the matter responsible for the energy loss should not have been
already detected. The natural candidate for this matter would be the dark
matter
in our galactic halo, or the cosmological dark matter.

The first part is achieved by assuming that the neutron has an initial energy
of
at least $ 10^{22}$ eV so that during its mean lifetime of $ \tau_{L} = 10^3$
sec it would travel a distance of $ c \gamma 10^3$ sec $= 100$ Mpc.  There are
two issues here, the identity of the particles and their energy source.
Regarding the first point we have already noted that once protons are produced
in any electron rich medium, neutrons with essentially the same energy can be
readily obtained via inverse beta decay.  In fact very high energy neutrons
might have already been observed$^{26}$ 
coming from our galactic
center, and it has been estimated that particles above the GZK energy coming
along the line of sight from CenA (about 3.4 Mpc away), are much more likely to
be neutrons than protons$^{27}$. 
Theoretical estimations also indicate
that neutrons should be a nontrivial fraction of the cosmic rays around the GZK
cutoff for nearby sources$^{28}$. 
Regarding the source of the very
energetic particles, we point out that nobody knows exactly what is the
mechanism producing the energetic protons, neutrons or other particles with
energies of the order of $10^{20} eV$.  The diffusive shock acceleration
mechanism (DSAM), is currently the standard theory of CR acceleration, but when
applied to supernova, it can acclerate particles up to $ 10^{17}$ eV. So, for
UHECR beyond $10^{17}$ eV, one has to invoke shocks on larger scale, for
example
in AGNs and radio-galaxies.  In fact, hot spots in Fanaroff-Riley type II radio
galaxies$^{29, 30, 31}$, 
about 100 Mpc away, and "Dead Quasars"$^{32}$ 
have been estimated to reach the $10^{21} eV$ mark, also Blazar jets$^{33}$  
have been estimated to be potential sources of protons with energies well 
above the $10^{20} eV$, 
while a recent proposal based on
the so called "Plasma Wake Field Acceleration Mechanism", seems to be able, in
principle to accelerate particles to any energy$^{34}$. 
As, we noted above,
our model requires sources of about $10^{22}$ eV neutrons, which, given the
above
examples and the astrophysical uncertainties, can not be excluded. As we have
already acknowledged this scheme makes the astrophysical problem posed by the
existence of cosmic rays with energies of order $10^{20} eV$ even more dramatic
by requiring particles of energies of order $10^{22} eV$ but in light of the
problematic aspects of other alternatives this seems a relatively small price
to
pay.

The loss of energy occurs either along the whole distance from the source $
D_S$
or just on the last section $D_L$ corresponding to the traversing of the
galactic halo of the Milky Way which is about $600$ kpc.

Regarding the second requirement we note that the SuperKamiokande detector
relies on C\^erenkov radiation. The recoil velocity of a nucleon moving with
the
Earth due to its interaction with a dark matter particle at rest in the frame
of
the galaxy would be at most $400$ km/s, and this velocity is too low to lead to
C\^erenkov radiation.  Another check is that the interaction of celestial
bodies -like  that of Earth in its motion with the sun around the galactic
center, or
that of the galaxy itself in the intergalactic medium --with these dark
particles should not lead to  noticeable effects.  In fact a bound for this
effect can be easily estimated for any ordinary object of mass $\mu$ and linear
size $R_o$ as it travels through the dark matter gas with velocity $v$.  We
envision these dark matter particles as Weakly Interacting Massive Particles
(WIMPs) and analyze the issue from the rest frame of the dark matter gas.  It
is
easy to see that the change in energy $E_o$ of the body in one interaction with
one WIMP of mass $M$ initially at rest is about $\Delta E_o= -Mv^2$.  We can
evaluate the change per unit time in the object's kinetic energy by simple
considerations leading to the following estimate for the fractional energy loss
\begin{equation}\label{Slowdown}
  \frac{ \Delta E_o}{E_o} \approx -\frac{\rho_{DM}}{\rho_o} \frac{ v
    \Delta t}{R_0} \times Min(1, \sigma(\rho_o/m) R_o)
\end{equation}
Now we consider, for instance the energy loss of Earth as it travels through
the
galactic Halo during the four billion years of the Earth's existence. Using the
density of the Dark Matter Halo, $\rho_{DM} = 0.3\quad$GeV/cm$^3$ Ref.[35], 
we find $\frac{ \Delta E}{E} \approx 10^{-9}\times Min(1,\sigma(\rho_o/m)
R_o)$.
Thus we conclude that these considerations lead to no useful constraint.

Finally we must address the tight bounds on the cross section of nucleon-WIMP
interaction for those WIMPs that might traverse our laboratories on Earth.
These are obtained by looking at the scintillation of heavy nucleus recoiling
from collision with WIMPS that might constitute the galactic dark matter halo.
The best such bounds are of the order $10^{-42} cm^2$ for he nucleus-WIMP cross
section.  We transform these into bounds on the neutron-WIMP cross section as
follows: 
\begin{equation} 
\sigma_{Nucleus-WIMP} \approx A^{2/3} \sigma_{neutron-WIMP} 
\end{equation}
where $A$ is the baryon number of the nucleus which for instance, for Ge --the
element employed in the most sensitive experiments 
to date$^{36,37}$ 
-- is 122. Thus we find; 
\begin{equation}
\label{CSbound} \sigma_{neutron-WIMP}\leq 10^{-43} cm^2.  
\end{equation}
We next analyze process of the neutron energy loss in its interaction with the
dark matter particles.

\section{ Analysis of the Energy Loss}\label{energy_loss}
We foccus on the rate of energy loss of an extremely relativistic neutron with
energy $\gamma m$ traveling through a gas of dark matter particles of mass
$M\neq 0$ assumed to be essentially at rest in our local comoving frame or that
of our galaxy. The neutron energy loss in one collision depends only on its
scattering angle $\theta$ and in fact we have: $ (\gamma \gamma'
-1)+(M/m)(\gamma-\gamma')=\gamma\gamma'\beta\beta' \cos (\theta) $ where
$\gamma$ and $\gamma'$ are the Lorentz factors for the incoming and outgoing
neutrons while $\beta$ and $\beta'$ the corresponding velocities (we are using
$c=1$).  Considering this equation in the extreme cases of $\cos(\theta)=\pm1$
we find,
\begin{equation}\label{DELTAE}
  \Delta^{(1)} E/E=-2 \frac{(M/m)(\gamma -\gamma^{-1})}{(1 +2\gamma (M/m)
    +(M/m)^2)}.
\end{equation}
The condition for energy loss to be smooth is $\Delta \gamma/\gamma<<1$( i.e.,
that it might be described as a continuous process). This implies either
$M/m<<(2\gamma)^{-1} $or $M/m>>2\gamma$.  On the other hand, the requirement of
forward scattering requires $ M/m<\gamma^2$. For the situation at hand, the
values of  $\gamma$ will be in the range $10^{11}$  to  $10^{13}$ so these
constraints are satisfied if $ 10^{13} GeV\leq  M \leq 10^{22} GeV$.  In that
case the energy loss in
one collision can be estimated as: $\Delta^{(1)} E/E=-2 (m/M)\gamma$.  Then
\begin{equation}\label{DE3}
  \frac{dE}{dt} = \int \Delta^{(1)} E \frac{dN}{dt}
  \approx  \Delta^{(1)} E~ n_{gas}~ \sigma~ v_{rel},
\end{equation}
where $\frac{dN}{dt}$ is the number of collisions per unit time. The RHS of the
last part of the equation has been obtained by assuming that all the collisions
have the same energy loss and all the particles in the gas were initially at
rest. Using $\Delta^{(1)} E\approx -2 E^2/M$ we have $\frac{dE}{dt} \approx
(-2E^2/M) (\rho_{DM}/M) \sigma $, where we have set $v_{rel} =1$ and
$n_{gas}=\rho_{DM}/M$.

We carry our analysis assuming that, in general, 
${dE}/{dt} =- {\mathcal{C}} E^{n}$. 
Note that, as the energy of our neutron in the rest frame of the gas of WIMPS
is a Lorentz scalar, such form does not imply any breackdown of Spacial
Relativity. $ {\mathcal{C}}$ is a constant that depends on the various
parameters
of the model. We then have, ${dE}/{d\tau} =- (E/m){\mathcal{C}} E^{n}$.
Considering $n$ non-negative we have:
\begin{equation}\label{time}
  \frac{1}{E_f^n} - \frac{1}{E_i^n} = \frac{\mathcal{C}}{m}n\Delta\tau
\end{equation}
for $ n>0;$ or $E_f =E_ie^{-\frac{\mathcal{C}}{m}\Delta\tau}$ for $ n=0$.
Let us examine first the case $n>0$: Using $E_i \ge 100 E_f$, we find, 
${\mathcal{C}} \approx m/(n{\Delta\tau} E_f^n)$.  
We can express the energy loss in
terms of the distance traveled. The infinitesimal distance is given by $dx =
\beta dt \simeq \gamma d\tau = (E/m) \ d\tau$, where we used the fact $\beta
\simeq 1$. Therefore, $ dE/E^n= -{\mathcal{C}} dx $ and
\begin{equation} \label{distance} 
D = \frac{1}{{\mathcal{C}} (n-1)} (E_f^{(1-n)}
  -E_i^{(1-n)}) \approx \frac{E_f}{m}(\frac{n}{n-1})\Delta \tau.
\end{equation}
While for the case $n=1$ we find, $E_f/E_i= e^{-\mathcal{C}D}$.  Armed with
these results, we now proceed to a more detailed analysis of the two different
scenarios.

\section{ Study of Different Scenarios}\label{scenarios}
Consider first the scenario where the neutron's energy loss occurs along the
whole trip from the source to the Earth.  Here we need $ D > 100Mpc$ and
$\Delta
\tau <\tau_{L}=10^3$ sec, while $E_f =10^{11}$ GeV and $m=1$ GeV. Thus for
$n>0$, using Eq.(\ref{distance}), we obtain $n/(n-1) >10^2$. So we need to be
essentially in the $n= 1$ case:
Defining $a=\frac{\Delta \tau} {\tau_{L}}$, so $ a<1$, and combining $E_f/E_i=
e^{-\mathcal{C}D}$, with Eq.(\ref{time}) we find 
$ (m/E_f) (1-e^{-\mathcal{C}D})= {\mathcal{C}} a \tau_{L} $.  Thus defining $x=
\mathcal{C}D$
and using the known values for $ E_f, m, \tau_L$ and $D$ we find $1-e^{-x}=
10^{-2}ax$.  For $ a \sim 1$ we have $x\approx 10^2/a$.  We will have $n=1$ if
the cross section is essentially $ \sigma=yE^{-1}$ with $y$ some constant. Thus
${\mathcal{C}}=2y \rho_{DM}M^{-2}$.  For the dark matter density in the
intergalactic medium we take the value $\rho_{DM}=\alpha\rho_{Crit}$ with
$\alpha \approx 0.3$ as indicated by the latest cosmological
data$^{38}$. 
We use $\rho_{Crit}=h^2\times 10^{-5}$ GeV/cm$^3$ with $h$ the
standard characterization of the Hubble parameter. We thus obtain the
expression
for the cross section $ \sigma =( 2a h^2)^{-1} [ M^2/(E\times 10^{13}$GeV$) ]
\times 10^{-6}$ cm$^2$. This model requires then a cross section that at low
energies is extremely large. This would seem to rule out this particular
scenario.  This conclusion might be avoided if for some reason the type of
WIMPs that make up the intergalactic matter are not present within the galaxy.  After
all our galaxy is located within a halo made out of of Dark Matter, which could
be of a very different nature from the intergalactic dark matter (in fact one
of
them seems to clump very effectively while the other does not) and is
conceivable that the two different kinds might repel each other.  One more
option to avoid dismissing this scenario emerges when we note that  there is no
need to
assume that the particular behaviour that is needed at the energies prevalent
in
the cosmic ray neutron interaction with the WIMPs extrapolates all the way to
the very low energy regimes where the cross section would become huge.  One
could view the $1/E$ behaviour of the cross section in the high energy regime
as
the tail of some resonance, and the low energy regime relevant for the
Laboratory
dark matter searches, as representing the other side of the tail or more
generally assume two very  different energy dependences separated by some
intermediate threshold.  None of these possibilities seems to be very elegant
or economic, but we must remain open to such possibilities in part by recalling
that if the standard matter sector of particle physics exhibits such richness,
the dark sector might do likewise.

In the second scenario where the neutron does not loose any of its energy until
it reaches the galactic halo, we assume the cross section behaves as $\sigma=z
E^N$. Then Eq.(\ref{DE3}) becomes
\begin{equation}\label{DE4}
  \frac{dE}{dt}   \approx   -2(\rho_{DM}/M^2) z E^{N+2},
\end{equation}
so $n=N+2$ and ${\mathcal{C}}=2(\rho_{DM}/M^2)z$.  Now the condition on the
distance traveled becomes $600$ kpc $=10^{14}$sec $\times (\Delta \tau /\tau_L)
(n/{n-1})$ which leads to $(\Delta \tau/\tau_L)$ 
$({N+2}/{N+1})=0.6$ which is
comfortably compatible with $N \sim 1$ as is clear that the proper time spent
in
crossing the galactic halo $\Delta\tau$ can not be but a fraction of the
neutron
lifetime $\tau_L$. The expression for the constant is thus ${\mathcal{C}}=
1.6\times 10^{-3}$ sec$^{-1} m(n-1)^{-1}E_f^{-n}$.  Setting $\rho_{DM}=0.3$
GeV/cm$^3$ the value for the dark matter Galactic halo, we obtain $z=8\times
10^{-14}$ cm$^2 \frac{1}{N+1}(M/E_f)^2 E_f^{-N}$ and thus 
\begin{equation}
  \sigma =8\times 10^{-14} cm^2 (N+1)^{-1}(M/E_f)^2 (E/E_f)^{N}.  
\end{equation}
In order to compare this with the bound of Eq.(\ref{CSbound}) we must note that
the quantity $E$ in the above formulae refers to the energy of the neutron in
the rest frame of the WIMP.  For the case of the conditions in the laboratory
which lead to the bounds in question this energy is that of a $300 km/sec$
neutron interacting with a WIMP, which is essentially the neutron rest energy,
thus:
\begin{equation}
  \sigma =8\times
  10^{-14} cm^2(N+1)^{-1}(M/E_f)^2 (m/E_f)^{N} \\
  =\frac{8}{(N+1)}10^{-10 -11 N} cm^2 < 10 ^{-43} cm^2
\end{equation}
where we have used $M=10^{13}$ GeV , $E_f= 10^{11}$GeV and $m=1$GeV. It is
clear
that the bound can be satisfied for $N > 3$. Thus for instance, taking $N=4$ we
have $\Delta \tau/\tau_L = (6/5)\times  0.6 < 1$. 

We have questioned the reliability of the estimations of the neutron CMB
photons
interactions, and thus we will consider as the minimal reasonable estimate of
the loss of energy due to this interaction the result of a calculation based
only on the interaction of the magnetic moment of the neutron with photons
using
the phenomenological lagrangian proposed by Pauli: $ {\mathcal{L}}_{PAULI} =
-\frac{\mu_p}{2}{\bar{\psi}}\sigma_{\mu\nu}\psi F^{\mu\nu}$.  We are interested
in
the elastic scattering between the neutron and the photon, which to lowest
order
gives, $ \frac{dE}{dt} \simeq -10^{-106}\mu_p^4T^2 E^4$.  Using $\mu_p \simeq
4.8 \times 10^{-7}$ MeV$^{-1}$, $T = 2.348 \times 10^{-10}$ MeV, we obtain a
value for ${\mathcal{C}}\simeq 4 \times 10^{-145}$ GeV$^{-2}$.  The energy loss
for such neutron, while traversing 100 Mpc is obtained from: $3 D {\mathcal{C}}
=
(E_f^{(-3)} -E_i^{(-3)}) = 1.8\times 10^{-105} $GeV$^{-3} $.  Thus, for
$E_f=10^{11}$ GeV, and assuming this to be the only interaction of neutrons and
the CMB photons, the energy loss  is negligible on the scales  and regimes we
are interested on.
 Of course  it is reasonable to expect this to be only a lower bound and
 there is plenty of room for  an increased cross section for neutron photon
interaction but as we argued we can not be absolutely certain of its value
based on the theoretical tools at our disposal today and the indirect methods
for its estimation which in turn rely on phenomenological calculational schemes
 whose extension to the case at hand  that we are questioning.  Again the
reader
should keep in mind that 
 we do this only in he spirit of questioning, among VMD and SR, the least well
established of the two.

\section{Conclusions}\label{conclusions}

The GZK anomaly is the only existing evidence presented as an argument
for Lorentz Invariance Violation. As mentioned before, this is a very
problematic 
proposal. In fact there is, up 
to this point, no truly  congruent model for the violation of Lorentz
symmetry$^{10, 11, 12, 13, 14, 15}$.
In view of the highly problematic nature of the 
proposals for  dealing with the GZK anomaly  on the basis of questioning of the
 Lorentz Invariance of physics, we have
considered   an alternative that  is based on the questioning of less  well
established principles.

We have briefly explored  here a  couple of  alternative 
scenarios, which have  the
advantage of not requiring any new  form of matter beyond the non-baryonic dark
matter  whose existence
is  required to address cosmological and astrophysical issues.  The
model does indeed  aggravate the requirements  on the source of high energy
particles beyond the levels required   in the context of  the observed cosmic
rays.  We just reiterate that this is already, and independently of this model,
a problem for astrophysics. It is clear the  need for a  thorough investigation
of the details of the proposals  mentioned 
at the beginning, and the search for   new alternative mechanisms
to deal with the  
astrophysical aspect of problem.
However the only  real new fundamental physics
that this scenario requires is an extremely weak
interaction of nucleons (baryons) with Dark Matter. One also  must be willing
to
call into question of phenomenological models of hadron-photon
interaction,  whose  extrapolation to
neutrons  has so far relied only on very indirect experimental studies.
This  aspects of our proposal have the added
attractive feature of making it  suitable for experimental
exploration within the context of dark matter searches, and more readily in
the experimental studies  of neutron-photon interactions and  their
confrontation with models such as VMD. In fact the nucleon-photon
interaction
in the energy regime of interest is, according to this analysis the
most pressing issue to  clarify the prospects of this kind of
solution to the GZK puzzle.
Our main point is that before considering the breakdown of something so
fundamental and well tested as Lorentz Invariance, and given the difficulties
that such program has
 encountered so far$^{10, 11, 12, 13, 14, 15}$,
 it seems worthwhile to
explore other options even if they involve relatively radical assumptions.  In
this respect we must keep in mind  that the degree of radicallity of such
assumptions must be measured relative to that of the proposals to do away with
special relativity. In this article we have shown that there is room for
"relatively  simple" explanations for the GZK enigma,  when considering that
the UHECR
might include neutrons. This  seems to be  an interesting  
option, that calls for further experimental exploration.

\section{ Acknowledgments}

We acknowledge very useful comments from Tsvi Piran.
This work was supported in part by CONACYT project 43914-F and DGAPA-
UNAM projects IN 108103 and IN 119405.

\end{document}